\documentclass[anonymous=false,sigchi]{acmart}

\usepackage{balance}       
\usepackage{graphics}      
\usepackage{subfig}
\usepackage{enumitem}

\def\BibTeX{{\rm B\kern-.05em{\sc i\kern-.025em b}\kern-.08emT\kern-.1667em\lower.7ex\hbox{E}\kern-.125emX}}
    
\copyrightyear{2018}
\acmYear{2018}
\setcopyright{acmlicensed}
\acmConference[Woodstock '18]{Woodstock '18: ACM Symposium on Neural Gaze Detection}{June 03--05, 2018}{Woodstock, NY}
\acmBooktitle{Woodstock '18: ACM Symposium on Neural Gaze Detection, June 03--05, 2018, Woodstock, NY}
\acmPrice{15.00}
\acmDOI{10.1145/1122445.1122456}
\acmISBN{978-1-4503-9999-9/18/06}

\begin{document}

%
\title[Tiger: Wearable Glasses for the 20-20-20 Rule]{Tiger: Wearable Glasses for the 20-20-20 Rule to Alleviate Computer Vision Syndrome}

%

\author{Chulhong Min}
\affiliation{
    \institution{Nokia Bell Labs, UK}
    }
\email{chulhong.min@nokia-bell-labs.com}

\author{Euihyeok Lee}
\affiliation{%
  \institution{KOREATECH, Republic of Korea}
    }
\email{euihyeok.lee@misl.koreatech.ac.kr}

\author{Souneil Park}
\affiliation{
    \institution{Telefonica Research, Spain}
    }
\email{souneil.park@telefonica.com}

\author{Seungwoo Kang}
\authornote{Corresponding author}
\affiliation{
    \institution{KOREATECH, Republic of Korea}
    }
\email{swkang@koreatech.ac.kr}

%
\renewcommand{\shortauthors}{Trovato and Tobin, et al.}

%

\begin{abstract}

We propose Tiger, an eyewear system for helping users follow the 20-20-20 rule to alleviate the Computer Vision Syndrome symptoms. It monitors user's screen viewing activities and provides real-time feedback to help users  follow the rule. For accurate screen viewing detection, we devise a light-weight multi-sensory fusion approach with three sensing modalities, color, IMU, and lidar. We also design the real-time feedback to effectively lead users to follow the rule. Our evaluation shows that Tiger accurately detects screen viewing events, and is robust to the differences in screen types, contents, and ambient light. Our user study shows positive perception of Tiger regarding its usefulness, acceptance, and real-time feedback.

\end{abstract}

%
%

\ccsdesc[500]{Human-centered computing~Ubiquitous and mobile computing systems and tools}

%
\keywords{Computer Vision Syndrome; 20-20-20 rule; Eyewear}
%

\settopmatter{printacmref=false}
%

\copyrightyear{2019}
\acmYear{2019}
\setcopyright{acmcopyright}
\acmConference[MobileHCI '19]{21st International Conference on Human-Computer Interaction with Mobile Devices and Services}{October 1--4, 2019}{Taipei, Taiwan} \acmBooktitle{21st International Conference on Human-Computer Interaction with Mobile Devices and Services (MobileHCI '19), October 1--4, 2019, Taipei, Taiwan} \acmPrice{15.00}
\acmDOI{10.1145/3338286.3340117} \acmISBN{978-1-4503-6825-4/19/10}

\maketitle

\section{Introduction}

Digital screens are ubiquitous and indispensable in our lives, but they are a double-edged sword. They benefit our productivity, entertainment, and information access; however, excessive use of them hurts our eyes. The prolonged use of digital screens causes various symptoms such as eyestrain, dry eyes, and blurred vision, referred to as Computer Vision Syndrome (CVS)~\cite{aoacvs, yan2008computer}. In addition, CVS symptoms also decrease the work productivity, thereby having a significant economic impact \cite{mark2011copmuter}. According to the report by the Vision Council \cite{2016digital}, nearly 90\% of Americans use digital devices for two or more hours each day. 65\% of Americans experience CVS symptoms. Our survey with 131 participants shows that they already experience various symptoms but they do not know the desirable practice of eye rest correctly.

To mitigate CVS symptoms, some previous works \cite{crnovrsanin2014stimulating, dementyev2017dualblink} attempt to stimulate eye blink. They provide intervention when detecting a decrease of blink rate while users view a screen. Although blinking prevents dry eyes, it is a partial solution to the CVS symptoms. The eyes use significantly more muscles when focusing on objects at a near distance, and continual screen usage easily causes eye strain \cite{yan2008computer}.

We propose Tiger\footnote{Tiger is an abbreviation of "TIme to Give your Eyes a Rest".}, an eyewear system to help users follow the 20-20-20 rule in their daily lives. The rule suggests to take a 20-second break to view objects of 20 feet away every 20 minutes of screen use \cite{aoacvs, aaoeyestrain, cao202020rule, 2016digital, yan2008computer}. The rule is one of the recommended strategies to alleviate CVS since taking frequent breaks to look at faraway objects significantly relieves the symptoms. Tiger monitors users' screen viewing activities and provides real-time feedback to help users take necessary actions. More specifically, Tiger notifies users that they need to 1) take a short break if they are viewing a screen for more than 20 minutes, 2) see 20 feet away if they are looking at nearby objects during the break, and 3) return to their previous activity if the break reaches 20 seconds.

We believe that the form factor of eyeglasses has unique advantages in achieving the goal. Detection of screen usages alone could be approached through \textit{screen-centric observation}, i.e., monitoring users from the device being used. Typical examples of this approach are timer-based reminder applications \cite{eyecare, EyeLeo, eyerestnotification}. There are also recent techniques that use a front camera of smartphones \cite{ho2015eyeprotector} or analyze keyboard/touch interactions for more accurate detection. However, developing a system for providing effective guidance to the rule goes far beyond simple screen view detection. It should also detect whether a user looks at something at a long distance while taking a rest. Moreover, the screen usages should be monitored \textit{across} devices as modern users are often equipped with multiple heterogeneous devices. We believe an eyewear enables \textit{viewer-centric observation}, i.e., observing all what a user sees, which can fulfill the above requirements. 

To build Tiger as a standalone wearable system, we address a range of challenges. First, we develop an eyewear prototype that carefully combines necessary sensors and actuators for accurate screen view detection and effective real-time feedback. Second, we design a light-weight, multi-sensory fusion technique for screen viewing detection. Rather than relying on a power-hungry camera-based approach, we fuse three sensing modalities, color, IMU, and lidar, each of which recognizes color patterns of objects being seen, head movement, and viewing distance, respectively. Third, we design real-time feedback to effectively guide users to follow the rule. To devise perceptible and comfortable feedback, we consider a number of parameters, e.g., actuator placement, actuation intensity, and duration, and identify suitable parameter values. To the best of our knowledge, Tiger is the first wearable system for the 20-20-20 rule. 

Extensive experiments are conducted to evaluate various aspects of Tiger. First, we collect 800 minutes long data under diverse realistic scenarios. Our evaluation results show that Tiger accurately detects screen viewing events, and is robust to the differences in screen types, contents, and ambient light. Second, our user study with 10 participants shows the positive perception of Tiger about its usefulness and acceptance. The real-time feedback of Tiger is also easily perceivable and comfortable.
\section{Related Work}

\textbf{The 20-20-20 rule and applications. }
Many ophthalmic organizations such as American Optometric Association, American Academy of Ophthalmology, and Canadian Association of Optometrists suggest following the 20-20-20 rule in daily lives to alleviate CVS \cite{aoacvs,aaoeyestrain,cao202020rule}. There are mobile and desktop applications as well as browser extensions devised to help users follow the rule \cite{eyecare, EyeLeo, eyerestnotification}. However, they simply provide timer-based periodic notifications regardless of whether a user actually views a screen or not. They are not able to check if a user takes a break looking at something 20 feet away. Also, they burden users to install on every device they use.

\textbf{Applications and systems for eye health. }
As mentioned, a major CVS symptom is dry eyes due to the reduced blink rate \cite{yan2008computer}. Some previous works proposed solutions under different environments to detect eye blinks automatically and provide intervention when the blink rate drops \cite{crnovrsanin2014stimulating, han2012eyeguardian, dementyev2017dualblink}. EyeProtector measures the viewing distance to a mobile phone using a front camera and alerts when the distance is too short \cite{ho2015eyeprotector}. While we believe that these works are complementary to our work for a comprehensive CVS solution, we focus on the recommended practice for screen viewing and build a tailored system to guide users according to the 20-20-20 rule. 

\textbf{Eye tracking \& gaze-based interaction.}
A large body of work has been made on eye tracking and gaze-based interaction for both stationary and mobile devices \cite{duchowski2002breadth, khamis2018past, delamare2017designing}. However, their primary goal is to accurately identify the point of the gaze or recognize a motion of the eyes given a display area. For accurate tracing, they commonly employ a hardware device tied to the target display area, e.g., a head-mounted display \cite{cauchard2011visual} or a camera \cite{vaitukaitis2012eye}. Though it is possible to adapt the solutions of these works for implementing the 20-20-20 rule, an application of them would be confined to a specific target display. Also, continuous usage of a camera and real-time image processing would incur heavy burden to battery-powered devices.

There are a few initial works based on eyewear devices for gaze tracking \cite{mayberry2014ishadow,mayberry2015cider,zhang2014starts}. As they are making progress on building power efficient solutions, they can be potentially helpful for detecting general screen viewing activities.

\textbf{Screen use detection. }
Zhang et al. proposed a system to detect moments of screen watching during daily life activities based on first-person videos from a wearable camera \cite{zhang2018watching}. Wahl et al. proposed a method for screen use detection using a color sensor in smart eyeglasses \cite{wahl2017computer}. In this paper, we find out that the sole use of the color sensor can make false-positive errors and thus propose a multi-sensory fusion including an IMU and a distance measurement sensor to make Tiger more robust to various situations. In addition to screen use detection, we devise effective feedback and build a wearable system to help users follow the rule. 
\section{Motivation}

\subsection{Screen Use \& Eye Rest Survey}

One may argue that people would relax their eyes well if they experience CVS symptoms and are aware of the importance of the eye rest. However, our user survey contradicts such a speculation. The survey was conducted with 131 participants recruited online (Table \ref{table:motiv_demographic} shows the demographics).

\begin{table}[t]
    \centering
        \begin{tabular}{|p{0.2\linewidth}|p{0.75\linewidth}|}
            \hline
            Age & 10s (2), 20s (89), 30s (18), 40s (7), 50s (15) \\
            \hline
            Gender & Male (81), female (50)\\
            \hline
            Occupation & Student (46), office worker (24), professions (21), researchers (11), sales job (8), other (21) \\
            \hline
        \end{tabular}
    \caption{Demographics of survey participants.\label{table:motiv_demographic}}
    \vspace{-0.2in}
\end{table}

\textbf{They already experience CVS symptoms.} We first asked the participants which CVS symptoms they had. Surprisingly, 99 of them (76\%) reported that they already experienced eye strain. The next common ones were neck/shoulder/back pain (57\%), blurred vision (40\%), and dry eyes (38\%). These numbers could be higher than those of the general public since many of the participants primarily use computers in their job.

\begin{figure}[t]
 \centering
 \includegraphics[width=.7\linewidth]{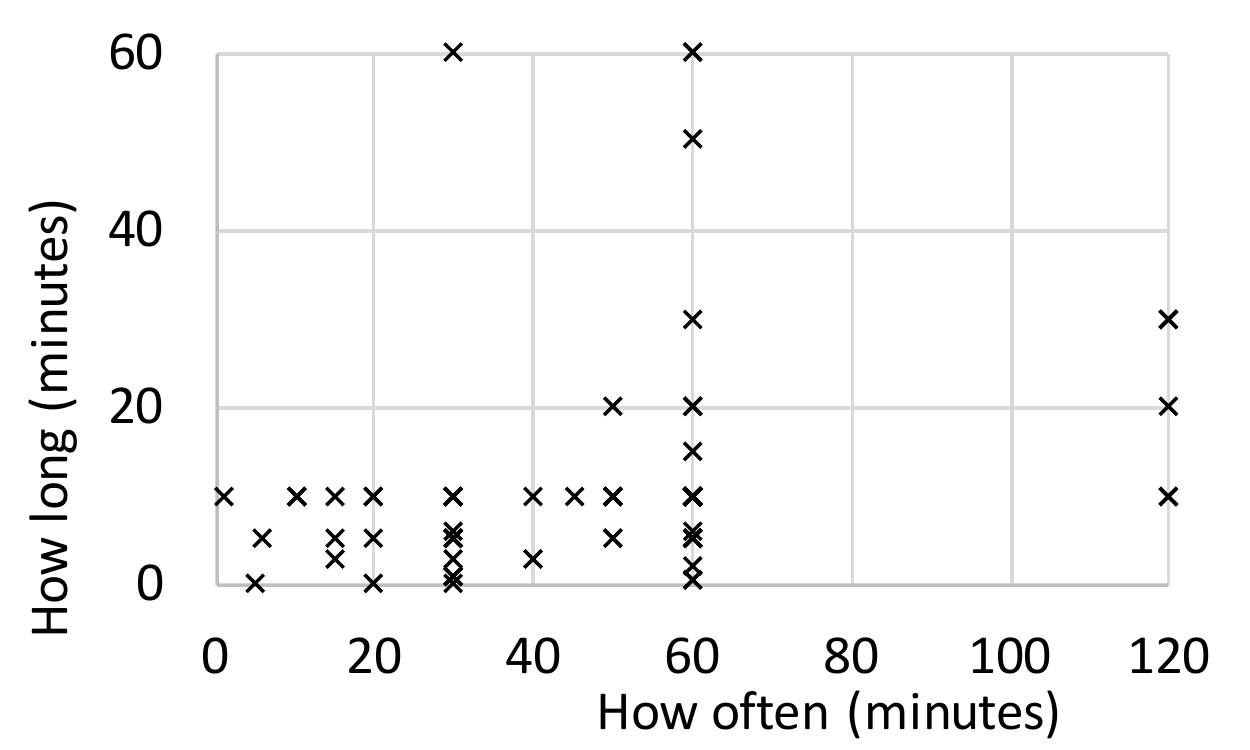}
 \vspace{-0.1in}
 \caption{How often/long should you give your eyes a rest?}\label{fig:motiv_resting_conv}
 \vspace{-0.2in}
\end{figure}

\textbf{They do not exactly know the recommended way of eye rest.} We further explored the participants' awareness of the desirable eye-resting practice. We asked what they think is a desirable rule for eye rest. Specifically, we asked how often and long they need to give their eyes a rest when they view screens. We asked them to write down the specific number and metric freely in either minutes or seconds. Figure \ref{fig:motiv_resting_conv} shows that users' awareness 
is very far from the 20-20-20 rule. While the recommended frequency of breaks is every 20 minutes, 87\% of them thought it is longer. As for the break duration, 97\% of them assumed that they need much longer time than 20 seconds; 87\% reported more than 5 minutes is needed. While there is no harm taking a longer break, we conjecture that such an assumption can make people feel reluctant to take breaks regularly.

\textbf{They do not follow the rule that they think is desirable.} 
We asked the participants how well they follow the rule that they think is desirable for regular rest in daily life. Out of 131 participants, only 4 and 10 respondents (3\% and 8\%) marked 'almost always' and 'often', respectively. 

\subsection{Limitations of Existing Apps for the Rule}
We conducted a preliminary study to investigate the limitations of existing applications for the 20-20-20 rule. We recruited 8 participants from a university campus (all males, 2 undergraduate and 6 graduate students). We asked them to install and use the applications on their computer and smartphone for a week; they were EyeLeo \cite{EyeLeo} for Windows PC, Breaks for Eyes \cite{Breaks} for Mac, Eyecare 20 20 20 \cite{eyecare} for iOS and Android smartphones. The actual use period varied from 5 to 7 days depending on the participants. After a week, we conducted group interviews with them, one with 5 participants and the other with 3. Each interview lasted for an hour.
Our key interview questions were: \textit{"How well did you follow the rule during the period of using the applications?", "If it was not successful, what was the main reason?", "What is your overall evaluation of the applications?"} The interviews were recorded and transcribed. Two researchers analyzed the transcripts individually, and discussed the main findings until they reached a consensus.

We summarize the limitations we found from the interviews. First, the applications often give inaccurate alarms about 20-minute screen viewing so that the users easily get annoyed. The applications do not track if the user actually views the screen or not. Especially we found that the mobile applications were simply timer-based and they gave notifications even when the screen was turned off. Such frequent false alarms discouraged the participants from using the applications. Second, they do not provide clear guides for the eye rest, especially for the viewing distance and the completion of the 20-second break. The desktop applications do show an alarm window to inform about the time for a break and suggest activities, such as looking out the window; however, the alarm simply disappears after 20 seconds. As a consequence, several participants often watched their screen again to check if 20 seconds has passed. In addition, the mobile applications simply send notifications in the same way as other applications do. The notifications were often ignored by the participants. Third, the current applications do not know if the users actually take a break. Moreover, they only monitor the screen use on the very device where they are installed. Some participants mentioned that they often unconsciously picked up and watched their smartphone during a break from a PC monitor.

\section{Tiger Overview}

Tiger is specifically crafted to help users follow the 20-20-20 rule in daily life by addressing the key requirements below. 

\textbf{Integrated monitoring of screen viewing.} As users can be exposed to many different kinds of screens (e.g., laptops, tablets, smartphones), Tiger aims to continuously monitor the screen viewing events in an integrated way across heterogeneous devices.

\textbf{Effective guidance of eye rest.} For adequate eye rest, Tiger should guide a user to look at something 20 feet away for at least 20 seconds. This leads to the following three sub-tasks: 1) detecting if a screen viewing event has continued for 20 minutes, 2) checking if the user sees 20 feet away during a break, 3) detecting if 20 seconds have passed.

\textbf{Non-distracting notification.} As observed in our preliminary study of existing apps, Tiger should avoid providing notifications via screen-equipped devices, e.g., smartphone, as it could turn a user\textquotesingle s attention into other digital contents.

Our basic approach is to build Tiger as an \textit{eyewear} system as it can directly track what a user sees. It allows continuous monitoring of screen viewing events for any devices, and measuring of the viewing distance during eye breaks. Also, the form factor of an eyewear gives advantages in terms of providing real-time feedback and helping users take appropriate actions at the right time. We also build Tiger as a \textit{standalone} system: it does not rely on external devices for sensing, data processing, or interacting with users. This enables Tiger to cope with various real-life situations where powerful devices such as smartphones are unavailable. As notifications are also sent directly from Tiger, users can avoid the possibility of being caught on another screen device.
\section{Tiger Implementation}

\begin{figure}[t]
    \centering
    \includegraphics[width=0.3\textwidth]{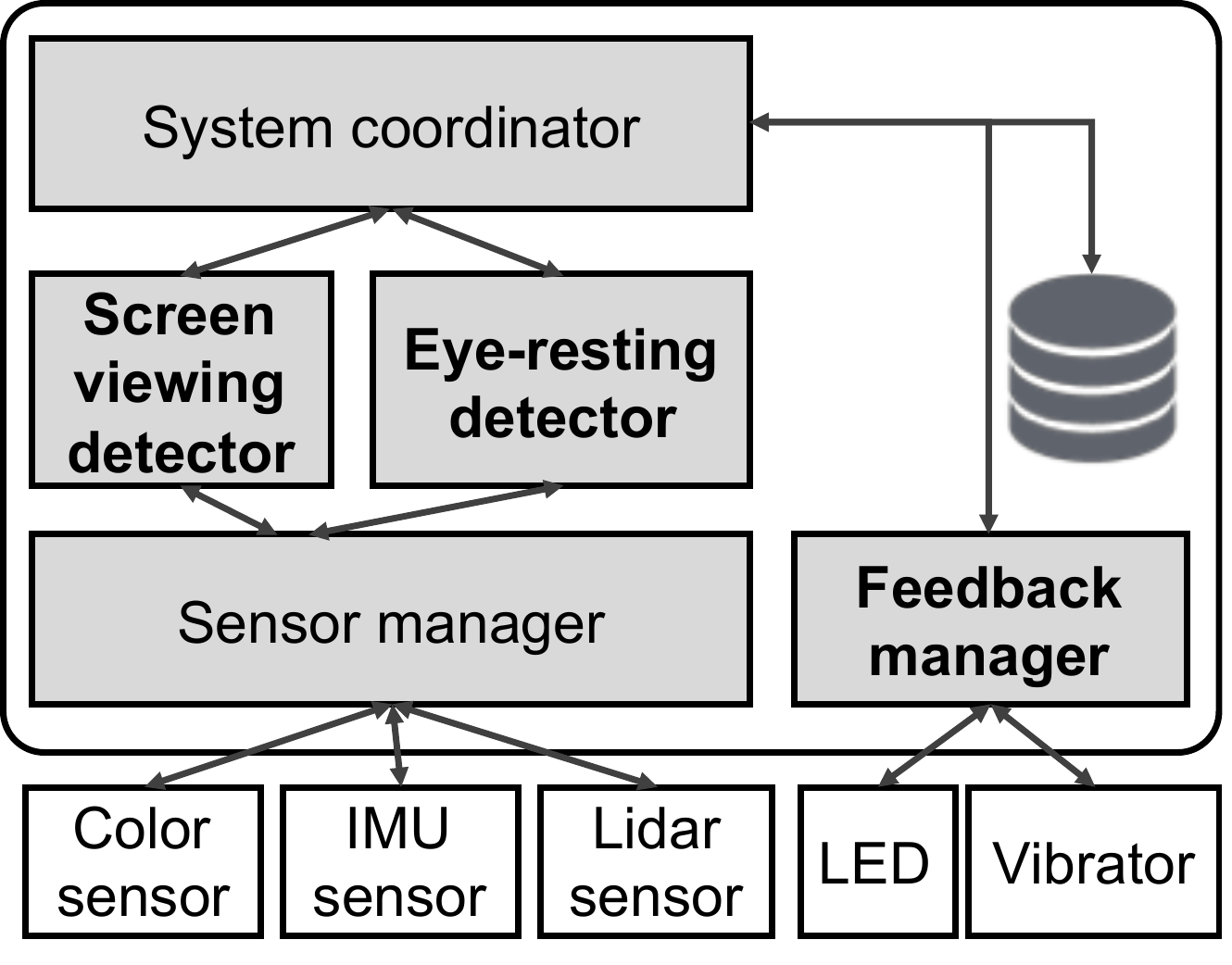}
    \vspace{-0.15in}
        \caption{Tiger architecture\label{fig:system_architecture}}
    \vspace{-0.1in}
\end{figure}

\begin{figure}[t]
    \centering
    \includegraphics[width=0.3\textwidth]{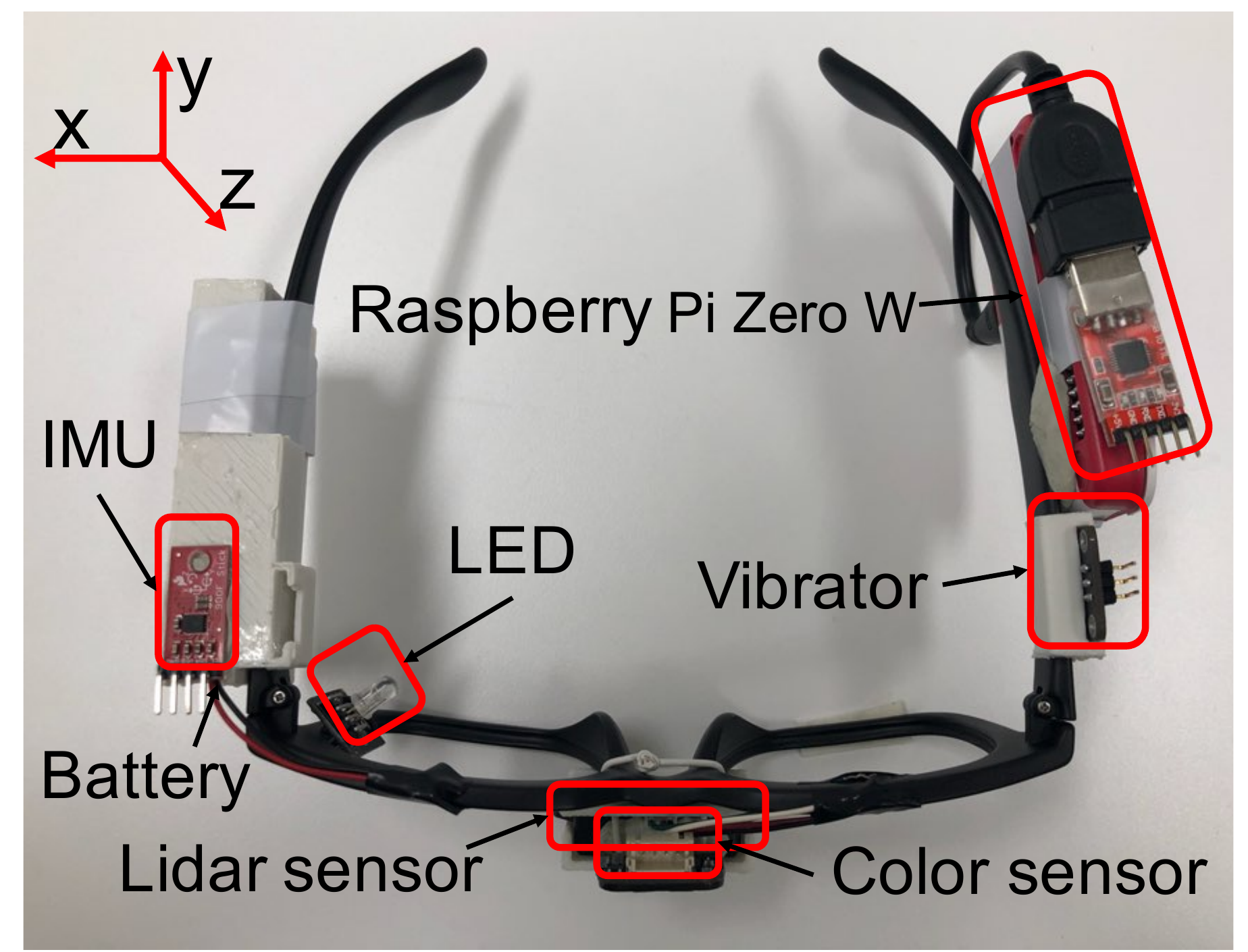}
    \vspace{-0.1in}
    \caption{H/W prototype\label{fig:hw}}
    \vspace{-0.1in}
\end{figure}

Development of Tiger involves diverse challenges including accurate screen viewing detection, non-obtrusive feedback design, and hardware fabrication. Moreover, the development requires a careful exploration of a design space, where the selection of sensors and subsequent data processing methods have to be made with respect to the specific requirements of our application. Our prototype is a result of examining a wide range of design options including sensor selection (e.g., camera, color, IR), their placement, and parameter optimization, etc. Various considerations such as privacy, power consumption, limited space of the eyewear frame, user comfort, were taken into account when making the decisions. We elaborate on the overall development and also discuss possible design alternatives in the following sections.

Figure \ref{fig:system_architecture} shows the system architecture of Tiger, whose key components are \textit{screen viewing detector}, \textit{eye-resting detector}, and \textit{feedback manager}. Figure \ref{fig:hw} shows the current hardware prototype \cite{lee2018towards}. As mentioned, all the components are mounted on an eyewear frame. The housings that hold all the sensors and the processing unit are custom-designed and 3D printed. The selected sensors for screen viewing detection are TCS34725 RGB color sensor, a SparkFun 9 DoF (Degree of Freedom) IMU sensor stick, and a TFMini micro LiDAR module\footnote{The module does not use laser light for ranging, but infrared based on ToF. Since it is marketed under the name "LiDAR", we use the term, lidar sensor for it in this paper.}. For the processing unit, we use Raspberry Pi Zero W with a 1000 mAh battery. 

Note that we use off-the-shelf sensors, actuators, and a glasses frame for fast prototyping. The main goal of this paper is to conduct a proof-of-concept study on the feasibility of eyeglasses device for the 20-20-20 rule. We believe a compact and well-integrated device with optimized size and weight can be developed with custom-designed processing units and hardware components, which is beyond the scope of the current work.

\subsection{Screen Viewing Detection Through Sensor Fusion~\label{sec:screen_viewing}}

We adopt a combination of three types of sensors: (a) RGB color sensor to sense the objects being seen, (b) IMU to sense head movements, and (c) lidar sensor to measure the viewing distance. While relying on a single type among them is prone to false detection, a proper combination can overcome the limitations of each type as the sensors offer complementary functions. We briefly discuss the rationale behind the choice of each sensor, and describe the combination method. 

\subsubsection{Color sensor - object being seen~\label{sec:detection_color}}

\begin{figure}[t]
    \centering
    \vspace{-0.20in}
    \mbox{
        \subfloat[Video on desktop\label{fig:impl_color_video}]{\includegraphics[width=0.23\textwidth]{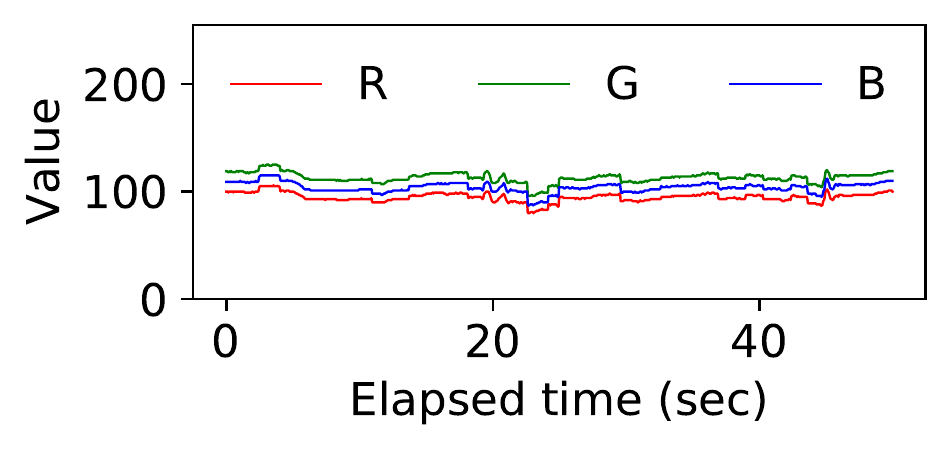}}
        \subfloat[Reading a book\label{fig:impl_color_book}]{\includegraphics[width=0.23\textwidth]{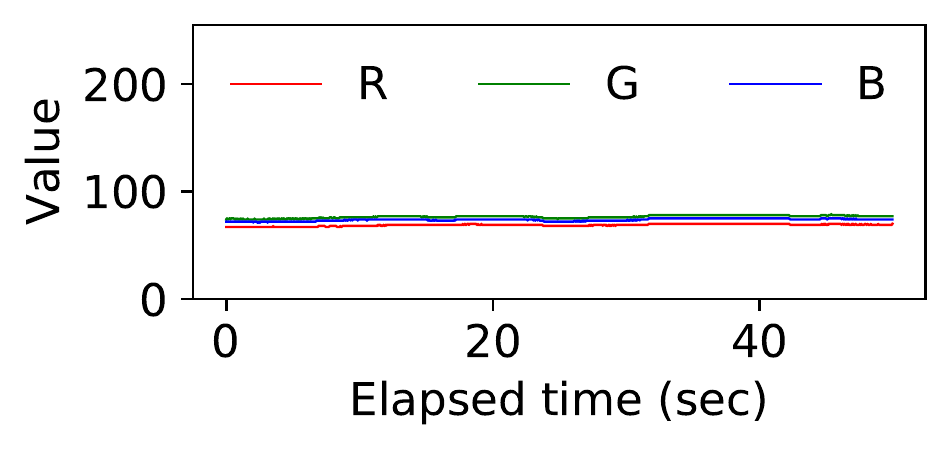}}
        }
    \vspace{-0.15in}
    \caption{RGB data of color sensor from P$_C$~\label{fig:impl_color}}
    \vspace{-0.15in}
\end{figure}

A key idea of using color sensor (RGB) is to leverage the speed of changes in the objects being seen. In our lab study, we obtained two insights. First, when users view a screen, they tend to stay still. The view and the objects being seen do not change much. Second, the changes mostly happen in the contents on the screen, and they change relatively faster than non-screen objects being seen, e.g., when reading a book. Figure \ref{fig:impl_color_video} and \ref{fig:impl_color_book} show the raw data of the color sensor in two example situations, when a user was watching a video on a desktop and when a user was reading, respectively; we used the traces obtained from the experiments in Section \ref{sec:eval_viewing}.

We choose a TCS34725 RGB color sensor that has red, green, blue, and clear light sensing elements. The same sensor was also used in Wahl et al.'s work \cite{wahl2017computer} for screen use detection, where they extracted time-domain features based on the relative ratios across the color channels and used SVM for the usage detection. We further optimize the screen viewing detection in two aspects. First, we use a different color scheme and propose new features for the color sensor. Second, more importantly, we additionally adopt IMU and lidar to achieve higher accuracy even in non-screen situations.

\subsubsection{IMU sensor - head movement~\label{sec:detection_imu}}

\begin{figure}[t]
    \centering
    \vspace{-0.05in}
    \mbox{
        \subfloat[Web surfing on laptop\label{fig:impl_imu_laptop}]{\includegraphics[width=0.23\textwidth]{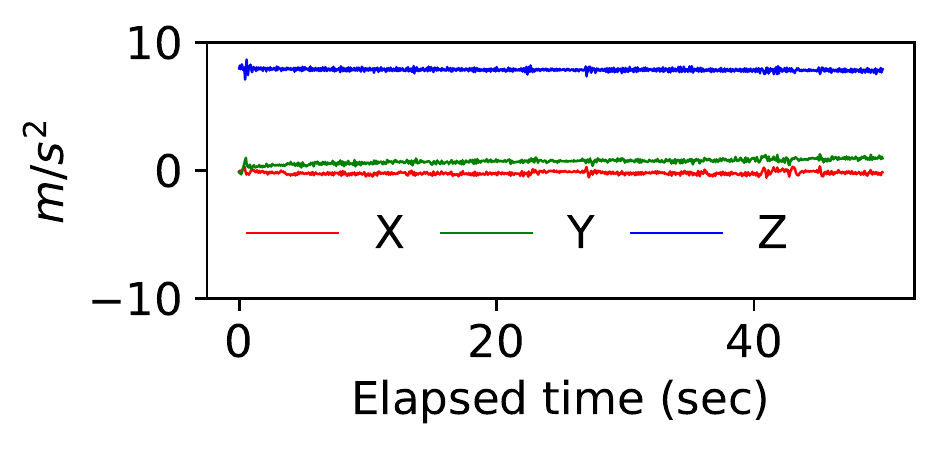}}
        \subfloat[Reading a book\label{fig:impl_imu_book}]{\includegraphics[width=0.23\textwidth]{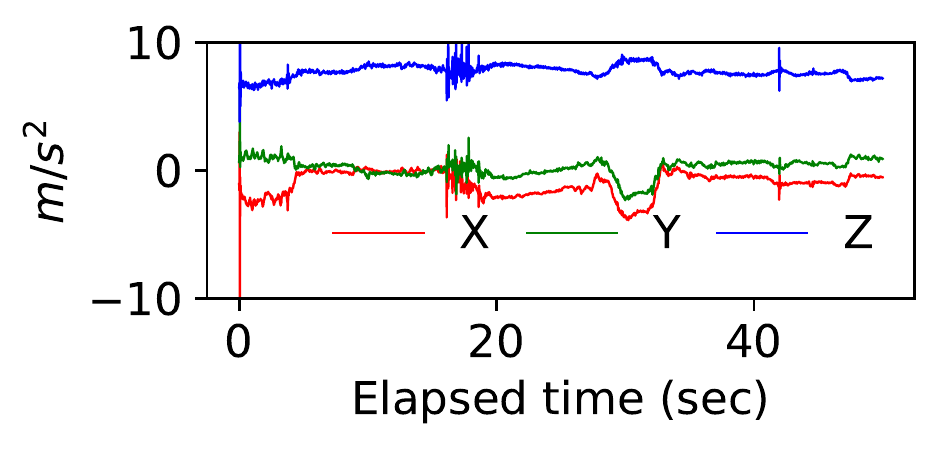}}
        }
    \vspace{-0.15in}
    \caption{Accelerometer data from P$_D$~\label{fig:impl_imu}}
    \vspace{-0.1in}
\end{figure}

The head movement can be a clear indicator of screen viewing activities. Users hardly move their head while watching a screen. The typical examples are working on a laptop and watching a video on a smartphone. Also, the head orientation would be a clue to detect screen viewing. People mostly view the phone and laptop while lowering the head and view the desktop screen while lowering or raising the head a little. On the contrary, people usually have relatively larger head movement when they do not view the screen even in the stationary situations. Figure \ref{fig:impl_imu} shows some examples (See Figure \ref{fig:hw} for the axis direction.) Figure \ref{fig:impl_imu_laptop} depicts the accelerometer traces when a user was surfing web on a laptop. Figure \ref{fig:impl_imu_book} is when a user was reading a book on a desk. The head movement was relatively bigger when a user read a book because he often looked around and nodded. Interestingly, the values of the Y-axis (user facing) both in Figure \ref{fig:impl_imu_laptop} and \ref{fig:impl_imu_book} are consistently higher  than those of the X-axis (horizontal) because people usually lower their head to see something.

\subsubsection{Lidar sensor - viewing distance~\label{sec:detection_lidar}}

\begin{figure}[t]
    \centering
    \vspace{-0.15in}
    \mbox{
        \subfloat[Web surfing on phone\label{fig:impl_lidar_smartphone}]{\includegraphics[width=0.23\textwidth]{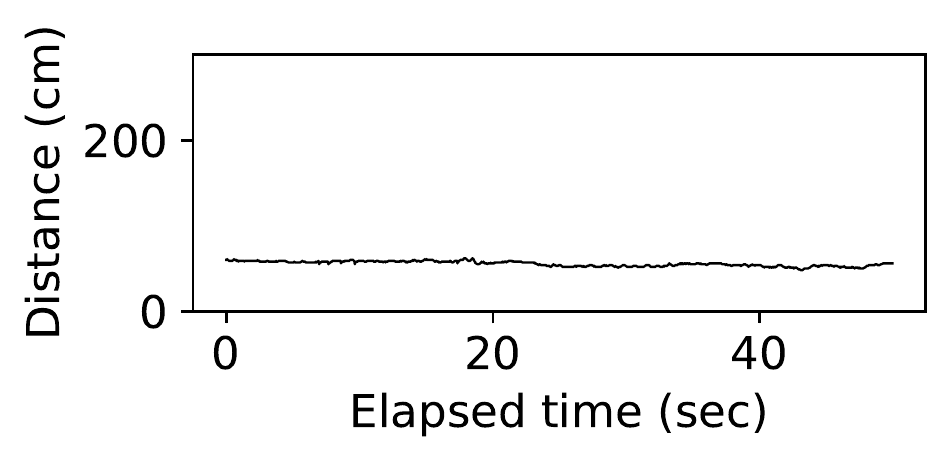}}
        \subfloat[Reading a book\label{fig:impl_lidar_rest}]{\includegraphics[width=0.23\textwidth]{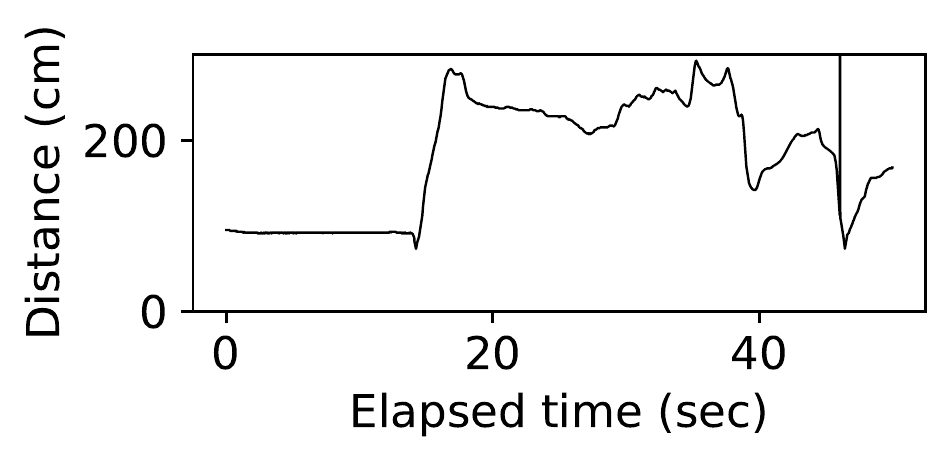}}
        }
    \vspace{-0.15in}
    \caption{Lidar sensor data from P$_G$~\label{fig:impl_lidar}}
    \vspace{-0.15in}
\end{figure}

According to \cite{long2014visual}, people have a typical viewing distance for digital devices. For example, people usually view a smartphone, a laptop, and a desktop at 18 - 60cm away, 40 - 70cm away, and 50 - 80cm away, respectively. Although the viewing distance in such ranges does not guarantee a screen viewing activity, the distance out of these ranges is highly correlated with non-screen activities. Figure \ref{fig:impl_lidar_smartphone} and \ref{fig:impl_lidar_rest} show the distance trace when a user was watching a video on a smartphone and reading a book, respectively. The sudden increase in Figure \ref{fig:impl_lidar_rest} was made because the user titled his head while reading a book. The glasses pointed to outside the book as a result, but he kept reading the book by lowering his eyes.

While the lidar sensor provides distance measurements, it is not trivial to obtain an accurate measurement. A slight mismatch of the lidar's orientation and that of the eye could result in a large error. After examining several placements and angles on our prototype, we decided to place the sensor on the bridge of the glasses. The accuracy was examined as the following. We test eight cases including short to relatively long distances, with 15 participants (10 males and 5 females) ranging from age 23 to 26 (mean: 24.6). We asked them to wear the Tiger prototype and view screen devices (a smartphone, a laptop, and a desktop screen). We also asked them to see five spots pointed by a laser distance meter (SWISS MILITARY SML-60M) in the lab, one at a time. The distance measured by the laser distance meter was compared with that of Tiger. Table \ref{table:distance_measure} shows the results. While the error ranges from 1.5 cm to 20.8 cm, we note that the errors are smaller for shorter distances. This error trend suits our application context well as the system does not have to be sensitive for distances larger than 20-feet.

\begin{table}
	\centering
	\begin{tabular}{ |c|c|c|c| } 
 	\hline
	  & \textbf{Distance meter} & \textbf{Tiger} & \textbf{Avg. error} \\
	\hline 
	 \textbf{Smartphone} & 29.6 & 31.2 & 1.6 \\ 
	\hline
	 \textbf{Laptop} & 50.2 & 48.4 & 1.8 \\
	\hline
	 \textbf{Desktop} & 60.8 & 52.8 & 8.0 \\ 
 	\hline
 	 \textbf{Spot1} & 125.9 & 128.4 & 1.5 \\
 	\hline
 	 \textbf{Spot2} & 261.1 & 288.0 & 16.9 \\
 	\hline
 	 \textbf{Spot3} & 540.5 & 561.3 & 20.8 \\
 	\hline
 	 \textbf{Spot4} & 696.3 & 711.7 & 15.4 \\
 	\hline
 	 \textbf{Spot5} & 838.5 & 842.9 & 4.4 \\
 	\hline
	\end{tabular}
    \caption{Distance measurement, unit (cm)\label{table:distance_measure}}
    \vspace{-0.20in}
\end{table}

\subsubsection{Data processing pipeline~\label{sec:detection_parameter}}

\hfill

\textbf{Color:} We read the RGB values at the interval at 42 Hz and convert the color space into hue, saturation, and intensity (HSI) since HSI is more robust to ambient factors (e.g., shadows or light reflection) \cite{xiong2018color}. In addition to the HSI stream, two additional streams that capture the similarity of the HSI samples of each time window are generated. The first stream computes the delta between two consecutive HSI samples, i.e., a list of $distance(X_i, X_{i+1})$, where $distance()$ is a distance function and $X_i$ is \textit{i}th HSI sample. The second stream computes the distance between all pairs of the HSI samples in a window, i.e., a list of $distance(X_i, X_j)$, where $j > i$. We used Euclidean distance as a distance function. From each stream, we compute five time-domain features: mean, median, variance, range between $80th$ and $20th$ percentile, and root mean square. In total, 25 features are extracted from a window.

\textbf{IMU:} The pipeline reads the data from 3-axis accelerometer and 3-axis gyroscope (86 Hz). The stream is segmented in time windows, and time-domain and frequency-domain features widely used for IMU-based activity recognition \cite{figo2010preprocessing} are extracted. Except for the correlation features, we take the magnitude of the 3-axis sensor values for the feature computation. The same set of features are computed for the accelerometer and gyroscope separately and combined. From a window, 38 features (19 from each) are extracted in total.

\textbf{Lidar:} The pipeline reads the lidar data at the maximum rate (66 Hz). We extracted the same set of features used for IMU except correlation and cross-correlation features. This is because changes in viewing distances also reflect head movements to some extent. In total, 13 features are extracted. For the eye-resting detection, we use an average of the distance values every second.

\textbf{Fusion:} Resulting features are merged into a single stream and normalised to a unified scale. Then, principal component analysis (PCA) is applied to reduce the input dimensions. Support vector machine (SVM) is used for the final classifier. 

We tune the hyperparameters, time window size, number of features (i.e., the number of PCA components), and the SVM parameters. As for the time window, we observe that longer windows produce better accuracy than the shorter ones. However, longer windows can incur the delay of operation, e.g., delayed feedback, and processing overhead. Considering our application context and the performance trends, we set the window size to 5 seconds. As for the degree of PCA feature reduction, we set the number of reduced features to be 50 as larger number of features does not bring much improvement. The hyperparameters of SVM was configured through grid search (kernel=rbf, C=100, gamma=0.001).

\subsubsection{Design alternative~\label{sec:design_alternative}}


An obvious alternative design would be to leverage an outward-directed camera and object recognition techniques \cite{zhang2018watching}. However, Tiger does not take this approach since the camera incurs significant system overheads regarding energy and CPU for continuous monitoring. More importantly, camera recordings can be significant threats to privacy and socially inappropriate. 

Alternatives were also considered for distance measurement: an ultrasonic sensor, an IR sensor, and a lidar sensor. We chose the lidar sensor since the first two were limited in terms of measuring distance and angle; distances longer than 20 feet were not covered and their measuring angle was relatively wide and causing errors. 

\subsection{Feedback for Eye Rest~\label{sec:feedback}}

We considered the following points in the design of Tiger feedbacks. First, a feedback should be effective for users to recognize the notification well even while focusing on ongoing activities such as work or study. Second, it should not be uncomfortable. Third, it should support various modes of feedback so that users can distinguish different notifications.

\subsubsection{Feedback Modality~\label{sec:feedback_modality}}
Tiger employs two types of feedback modality, \textit{vibration} and \textit{light (LED)}. We do not choose auditory feedback because users can be in situations where auditory feedbacks are inappropriate, e.g., shared spaces or noisy places. To our knowledge, the two selected modalities have not been studied much for an eyewear. A few works study the effectiveness of directional haptic cues for navigation systems \cite{nukarinen2015delivering} and for feedback of gaze gestures \cite{rantala2014glasses}.

We thus conduct a pilot study specific to our context, to identify an effective configuration for each modality regarding perceptibility and comfortability. We mainly consider two parameters, vibration strength and LED position. Other parameters, e.g., vibration duration and LED intensity were set based on prior works \cite{saket2013designing, kaaresoja2005perception} and our internal tests. For the vibration strength, we test four options from weak (20) to strong (50). For the LED position, we test three options, i.e., top, middle, and bottom right side of the right rim. 

We asked 10 participants (the same participants of the first experiment in Section \ref{sec:eval}) to wear the glasses and perform three tasks, each of which lasts for 30 seconds. The tasks were 1) taking a rest, 2) typing keyboard of a desktop computer, and 3) using a smartphone. During a task, a feedback using the LED light or vibration was sent. After the task, we asked them to answer two 7-point Likert scale questions regarding the perceptibility of the feedback (1: never perceptible - 7: very well perceptible) and comfortability (1: very uncomfortable - 7:  very comfortable). 

\begin{figure}[t]
    \centering
    \includegraphics[width=0.5\textwidth]{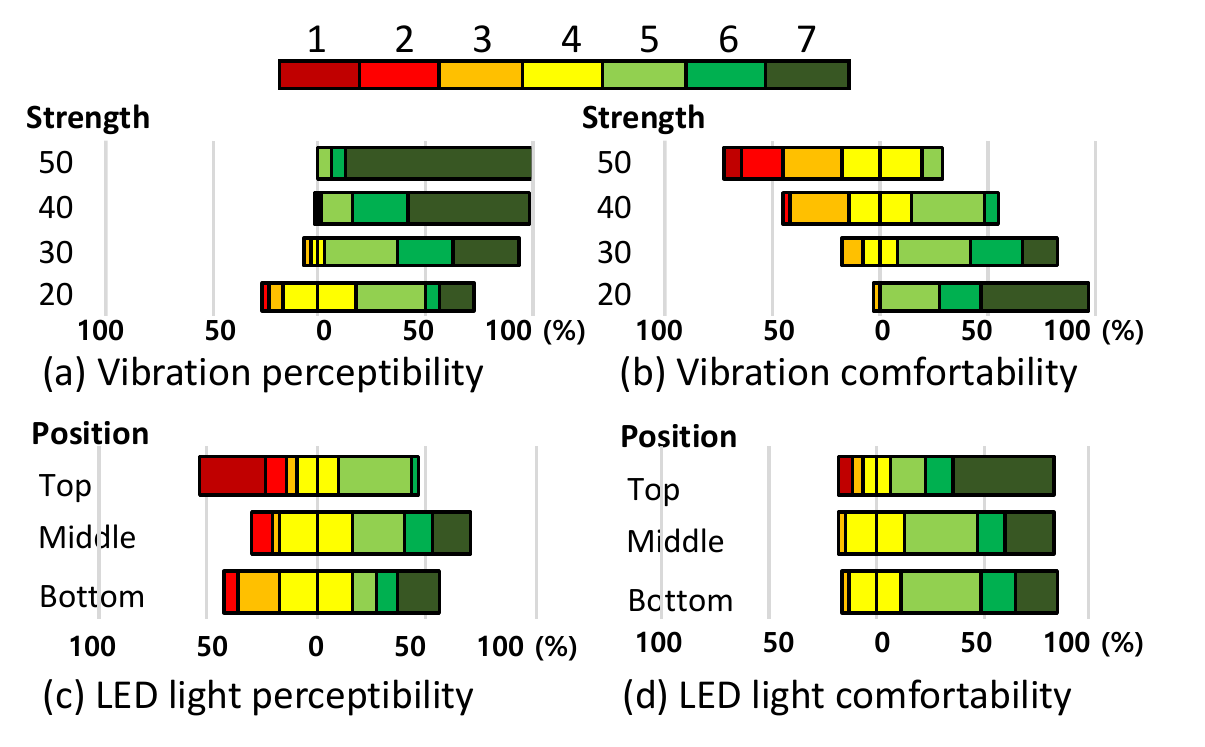}
    \vspace{-0.35in}
    \caption{Feedback modality test}
    \vspace{-0.2in}
    \label{fig:feedback_modality}
\end{figure}

Figure \ref{fig:feedback_modality} shows the distribution of ratings. Overall, the vibration is more perceptible but more uncomfortable than the LED. Regarding the vibration strength, the option 30 offers a good trade off between perceptibility and comfortability, hence we choose this option for alarming users to take a break. For the LED, all three options have similar comfortability. Thus, our choice was made based on the  perceptibility criteria, taking the option of the middle position. As for the vibrator, two positions were considered, the bridge and temples. We observed that putting it on the bridge makes the nose pads vibrate, making a user feel tickled. The left temple did not have such an issue and the subjects perceived the vibration well.

\subsubsection{Feedback Design}

Tiger sends three types of feedback: (1) 20 minutes of screen viewing, (2) 20 feet of viewing distance during a break, and (3) completion of 20 seconds of break. 

We believe the first feedback type (20 min. screen viewing) should be immediately noticeable and give a sense of urgency as it should lead a user to notice the situation and take a break. We thus use vibration, which is more perceptible than LED. The vibration feedback is designed to alter between a relatively long vibration (600 ms) and a short break (200 ms). The length of each state was made based on the previous study on the vibration alerts on smartphones \cite{saket2013designing}. 

As for the second feedback mode, we use the LED light only. The rationale is twofold. First, we give high priority to comfortability over perceptibility since a user stops viewing a screen anyway. Second, we can reduce the potential cognitive burden of users to remember and distinguish several different vibration patterns. We turn on the green LED when a user is seeing further than 20 feet (or longer) away. Otherwise the LED turns to red to warn that the viewing distance is not far enough. We set the LED to flash in a period of one second.

The third feedback type is simply for notifying that 20-seconds have passed so it is okay to go back to the screen. We believe this feedback does not have to be strong as it is not necessary to push users to go back to the screen. Thus, we use a weak vibration (20), altering between on and off (each 600 ms). This continues for 5 seconds and the LED light that was on during the break session is turned off. 

\section{Evaluation~\label{sec:eval}}

We perform extensive experiments to evaluate the following aspects. First, we examine the accuracy of the screen viewing detection under a range of realistic settings with different screen devices. Second, we conduct a user study to investigate the user-perceived implications of Tiger in terms of the acceptance and usefulness as well as the real-time feedback.

\subsection{Screen Viewing Detection~\label{sec:eval_viewing}}

We evaluate the screen viewing detection of Tiger at two levels of temporal granularity: 1) detection for a short time window (i.e., 5-seconds) which is the minimum unit of the detection in our experiment, 2) detection of  20-minute screen-viewing sessions. We first describe our data collection method, and elaborate on each evaluation in more detail below. 

\subsubsection{Screen View \& Non-Screen View data collection}

\hfill

\textbf{Participants. } We collect data of screen views and also non-screen views from 10 participants (male: 9, female: 1). They are recruited from a university campus. All of them are undergraduate students and their ages are between 24-27 (mean:25.3). Seven of them wear glasses. Their periods of wearing glasses are between 11-17 (mean:13) years. It is important to note that Tiger does not target eyeglasses users only, but also includes non-eyeglasses users because the main goal of the 20-20-20 rule is to prevent CVS. None of them had any prior knowledge of Tiger. Each participant was compensated with a gift card equivalent to USD 18. 

\textbf{Procedure. } Each participant was invited to the lab and asked to follow a series of scenarios while wearing Tiger. Each scenario is either a screen view or non-screen view session, where we vary a number of parameters to reflect diverse real situations. The parameters are the followings. 

\begin{itemize}[leftmargin=*]
\item Device: We considered three screen devices: a smartphone, a laptop, and a desktop screen. We asked the participants to use their own smartphone (2 iPhones and 8 Android phones) and provided them with a laptop (LG gram 15" Ultra-Slim Laptop with Full HD IPS Display) and a screen (LG 27" Class Full HD IPS LED Monitor) with a desktop. 

\item Content: We considered two types of contents on the screen, \textit{static} and \textit{dynamic}. For static contents, we asked the participants to freely surf the web: the main contents were articles on news and online communities, and comics. For dynamic ones, they played a game or watched a video.

\item Non-screen activities: We considered two types of activities for the non-screen view sessions, \textit{reading a book} and \textit{taking a rest}. For the former, the participants read a book in a natural position, which they chose freely. For the latter, they freely moved around the laboratory or had a chat with the researchers.

\item Ambient light: For both screen and non-screen view sessions, we considered two ambient lights, namely \textit{dark} and \textit{bright}, by turning off and on the light in the lab. The average ambient luminance were 100 and 300 lux, respectively. 
\end{itemize}

To sum up, each participant followed 16 scenarios (16 = $3 \text{ devices} \times 2 \text{ contents} \times 2 \text{ lights} + 2 \text{ non-screen activities} \times 2 \text{ lights}$). Each scenario lasted for 5 minutes, thus, 800 minute-long data were collected in total.

\subsubsection{Detection for 5-second windows}

We perform a binary classification, \textit{screen view} vs. \textit{non-screen view}, for all the 5-seconds segments of the entire collected data. A 10-fold cross-validation (10-fold) is conducted and the $F_1$ score, which is a harmonic mean of precision and recall, is used as our metric. 

\textbf{Overall performance.} The results show that Tiger detects non-screen/screen activities accurately. The average $F_1$ scores were 0.98 and 0.99 for the non-screen and screen views, respectively. The performance is slightly better for detecting screen views because the behaviors tend to stay stable when people are in front of a screen (e.g., object of focus, head movement, and viewing distance), hence similar sensing signals are generated. In contrast, there is more possibility of variation for the non-screen view activities. For example, one participant just sat on a chair during the whole 5 minutes to take a rest while some other participants moved around the laboratory and had a chat with the researchers. 

\textbf{Effect of parameters.} Recall that we varied a number of parameters, i.e., contents, ambient light, and devices. We breakdown the performance according to each parameter, however, observe that they hardly have an impact on the performance. The results for the screen content and ambient light show high $F_1$ scores, all above 0.95. For the various conditions of used devices, the average $F_1$ scores are between 0.96 and 0.98. This indicates the robustness of Tiger to different conditions in our scenarios.

We also conduct additional analyses considering the effect of sensor availability situations and training methods. 

\begin{figure}[t]
    \centering
    \includegraphics[width=0.35\textwidth]{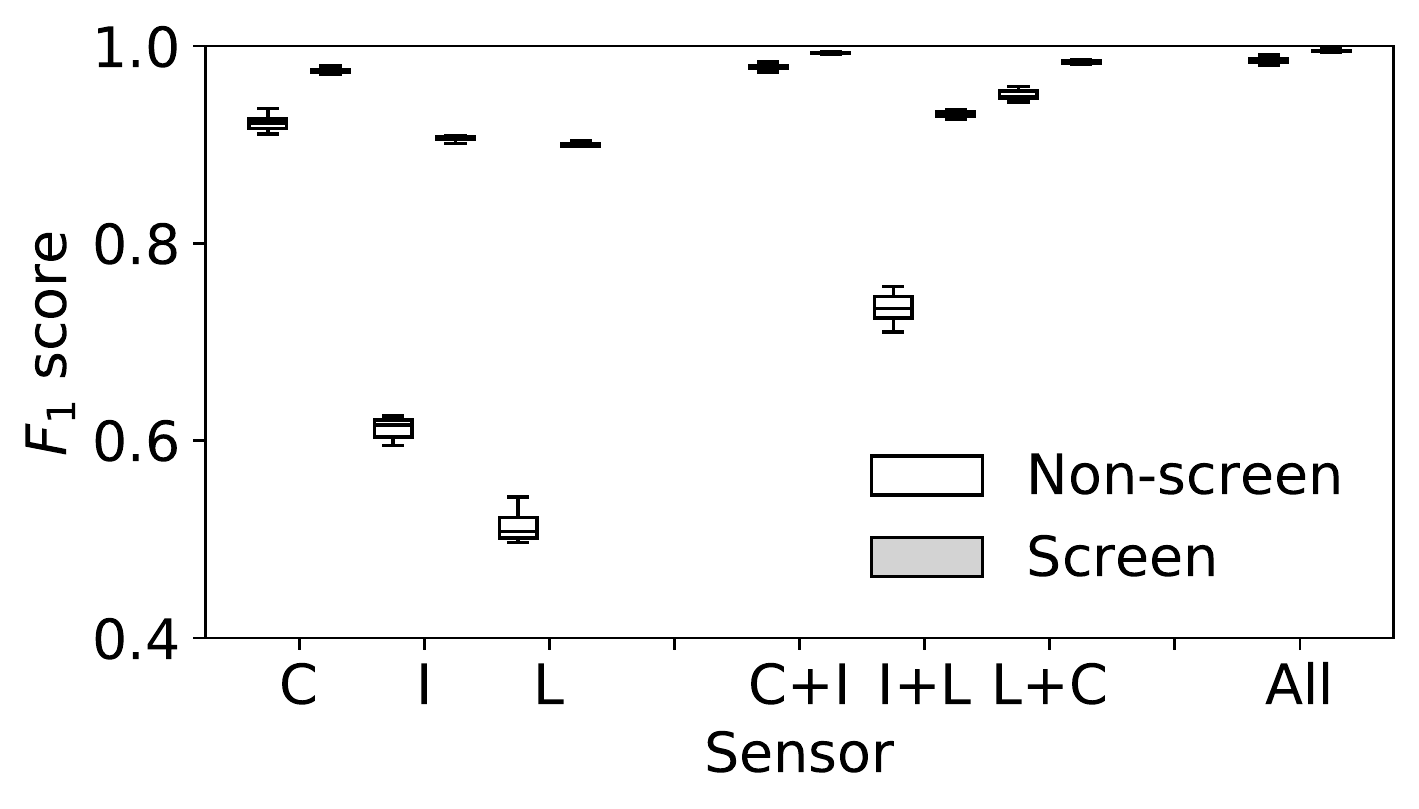}
    \vspace{-0.15in}
    \caption{Effect of Sensor\label{fig:eval_sensor}}
    \vspace{-0.2in}
\end{figure}

\textbf{Effect of sensor availability.} Assuming situations where certain sensors are not available, we investigate the performance achieved with different sensor combinations. Figure \ref{fig:eval_sensor} provides an overview; C, I, and L represent the color, IMU, and lidar sensor, respectively. In general, the detection of non-screen views is affected seriously when any of the three sensor is not available. We believe this is due to the diversity of possible behaviors during non-screen views in contrast to the regularity of behaviors during screen views. 

When a single sensor is used, the average $F_1$ scores for the non-screen views are 0.92 (color), 0.61 (IMU), and 0.51 (lidar), respectively. The lower performance of IMU and lidar is expected considering that head movement (IMU sensor) and viewing distance (lidar sensor) in some of non-screen activities are similar to those of screen activities, e.g., reading a book and viewing a laptop. When two sensors are used, the pair of the color sensor and IMU (0.98 and 0.99 of the non-screen and screen activities) outperforms the other pairs. Using all three sensors achieve the best result, while the difference in $F_1$ scores is not much significant compared to the pair of the color and IMU sensors. In case that energy efficiency is the first priority, using the pair only would be beneficial in terms of the power consumption. We believe Tiger can adopt a conditional sensing pipeline to achieve the maximum accuracy while saving the energy, e.g., using IMU and color sensors in the first stage and triggering the lidar sensor when needed.

\textbf{Effect of training methods.} While 10-fold is a widely used validation method, it is assuming that a model is trained with the data of all the users. We analyze the performance assuming two additional situations: first, Tiger has to adapt to a new user who hasn't been seen during training; second, Tiger has data only from a single user. As for the first situation, we analyze the performance through a leave-one-subject-out (LOSO) cross-validation, where a model is trained with all data except one participant and tested it for the data from the excluded participant. As for the second, we analyze the performance for the \emph{personalized} models, where we trained and tested the models separately for each participant (referred to as \textit{P$_A$} to \textit{P$_J$}). The model is trained with 60\% of the data of each participant, and tested it with the rest. 

Tiger shows a certain gap in the performance between the original 10-fold and LOSO, which implies the performance drops when Tiger is used by a new, unseen user. For LOSO, the average $F_1$ scores are reported as 0.79 for non-screen views and 0.93 for screen views. The main reason is the diversity of behavioral patterns across users during non-screen viewing activities as mentioned above. On the other hand, performance is again high for the personalized models. P$_A$'s average $F_1$ scores of the non-screen and screen views are 0.98 and 0.99, respectively. For the rest, all the average $F_1$ scores both for the non-screen and screen activities are reported above 0.99. In this regard, we can expect Tiger would operate well in deployment environments if it adopts an \emph{online learning} technique that collects sensor data on the fly and gradually trains a user-specific model.

\subsubsection{Detection of 20 minute Screen View Sessions}

Our target application ultimately requires the detection of the 20 minute screen view events. Accurate detection of the events is not trivial in real-life situations as people often take intermittent breaks and move their views out from a screen. If the break lasts for 20 seconds, the measurement of the session should start again. We conduct a separate analysis considering such situations. Tiger basically performs the detection by accumulating the results of the 5 second windows; a sequence of 240 screen view windows triggers the detection of the session. To deal with intermittent breaks, Tiger traces if a non-screen event lasts for 20 seconds. If four consecutive non-screen events are detected in a row, it assumes that a user has taken an eye rest and resets the tracking of the 20 minute session. Otherwise, the tracking of the session continues. 

Since it is extremely challenging to collect data of long-term screen viewing sessions under diverse conditions in a natural setting, we synthetically generated 20 minute screen viewing sessions from the data collected for the previous experiments. More specifically, for each participant, we generated 1,000 20-minute-long sessions where a session consists of three parts [screen-view, non-screen-view, screen-view]. The screen-view part of a session is randomly sampled from the 12 screen viewing scenarios, and the non-screen-view part is sampled from the 4 non-screen scenarios. We varied the duration of non-screen-view breaks from 5 to 35 seconds at the interval of 5 seconds. 

The results show that Tiger is robust to the intermittent breaks. The average $F_1$ score of classifying screen-view and non-screen-view sessions is 0.94 and 0.93, respectively. Table \ref{table:eval_20minutes} shows the detection accuracy for each duration of a break. The accuracy is very high (over 0.98) for the sessions with breaks less than 20 seconds. This is because it is unlikely that Tiger performs wrong detection for several 5-second windows in a row. As for the sessions with breaks around 20 seconds, the detection becomes more challenging since incorrect detection for a single 5-second window can lead to an incorrect classification of a 20-minute session. The accuracy drops when the break duration is exactly 20 seconds, but it increases again as the duration increases.

\begin{table}
    \centering
    \begin{tabular}{|p{2cm}|c|c|c|c|c|c|c|c|} 
     \hline
     Break (sec) & 5 & 10 & 15 & 20 & 25 & 30 & 35 \\
     \hline 
     Accuracy & 1.0 & 1.0 & 0.98 & 0.85 & 0.87 & 0.91 & 0.95\\
     \hline
    \end{tabular}
    \caption{Accuracy for 20 minute-long screen sessions\label{table:eval_20minutes}}
    \vspace{-0.3in}
\end{table}

\subsection{Tiger User Study~\label{sec:eval_userstudy}}

\subsubsection{Participants}
We recruited 10 participants (male: 6, female: 4) from a university campus. They are undergraduate students and their ages are between 23-26 (mean: 24.4). Most of them are heavy screen users; they thought they use screens for more than 5 hours a day on average. Five of them wear eyeglasses. Six of them have their own practice for eye health, e.g., trying to take a regular break, trying to see far away or do eye massage. Everyone voluntarily participated and was compensated with a gift card worth USD 22. 

\subsubsection{Tasks}
We considered two scenarios: one was to use a computer for work and the other was to freely spend time using screens. The user study comprises two tasks accordingly. For the first task, we gave a paper to the participants and asked them to type the text of the paper on a computer. To encourage them to concentrate on the task as if it were their actual work, we offered a prize for one who typed the most amount of text correctly; the prize was a gift card equivalent to USD 9 to a winner and a gift card equivalent to USD 4 to two runner-ups. The first task was about one-hour long, so they were guided by Tiger to follow the rule three times. For the second task, we asked the participants to do anything they want with using a screen, e.g., playing a game, watching a video, surfing the web. We let them use any screens they want among a desktop monitor, smartphone, and laptop. The second task also lasted for one hour. 

\subsubsection{Procedure}
The study consisted of three parts. First, we gave the participants a brief explanation about the study and the functionality of Tiger, and got written consent. Second, they performed the aforementioned tasks for two hours while wearing the Tiger prototype. Last, we conducted a questionnaire and a semi-structured interview individually. The interview was about the acceptance and usefulness of Tiger, and perception of the provided feedback.

\subsubsection{Results}

\hfill

\textbf{Acceptance and usefulness of Tiger.} 
Most of the participants showed their positive perception of Tiger. Nine of them reported that Tiger's feedback was useful to follow the 20-20-20 rule and eventually it would be helpful for their eye health. P4 said, \textit{"Usually when I focus on something, I can't pay attention to other things, but the vibration made me realize that I need to take a break and I could stop viewing the screen and try to see something far away. I think because of this my eyes don't feel strained much."} P7 mentioned, \textit{"Explicit notification allowed me to do an action appropriate for that situation. (...) I usually feel my eyes get dry when I view a monitor, but this time I didn't feel that much."} P9 agreed that it was useful to follow the rule, but showed her concern about potential side effect. She said, \textit{"(...) But I think LED is too close [to my eyes] and (...) it might be harmful to my eyes."} On the other hand, one participant (P6) commented that two hours is not enough to tell the usefulness.

When our participants were asked \textit{"If there is a commercial product that provides the functionality of Tiger, are you willing to buy and use it?"}, 7 of them responded positively. P1 stated, \textit{"I'm very interested in the health of my eyes (...). I usually try to take a break regularly, but it is hard. I am  willing to use a product like Tiger ..."} P2 mentioned, \textit{"If its price is reasonable, I will [use it]."} P9 said, \textit{"Usually I don't even know how much time has passed when I use a smartphone. If it gives me notification every 20 minutes, I can take a break even if it is short. So I will use it."} In contrast, three participants told that they would not use it. P4 showed her negative feeling about wearing eyeglasses itself.

\begin{figure}[t]
    \centering
    \includegraphics[width=0.5\textwidth]{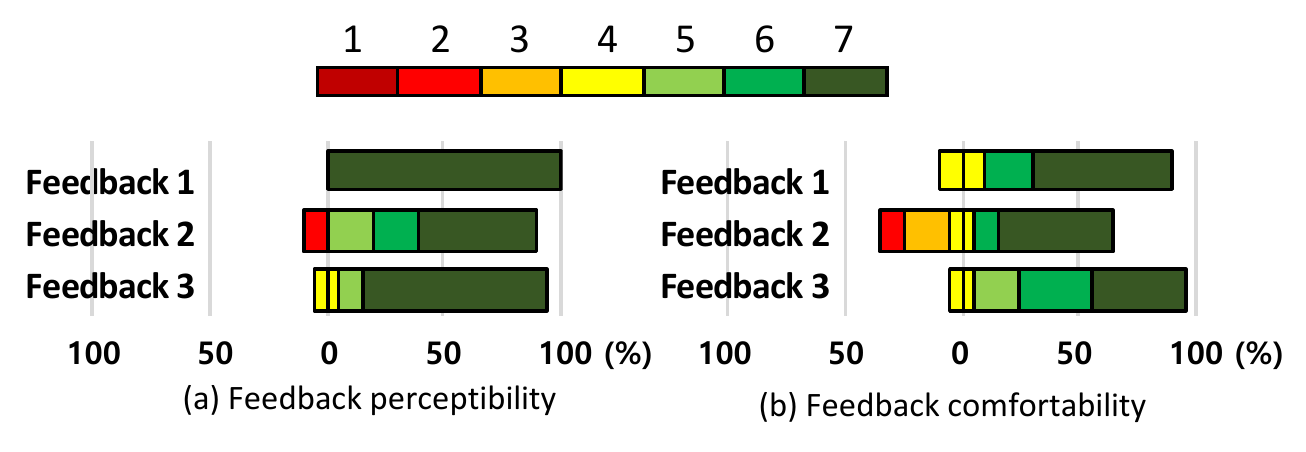}
    \vspace{-0.3in}
    \caption{Feedback score distribution.}
    \label{fig:feedback_rating}
    \vspace{-0.23in}
\end{figure}

\textbf{Assessment of the feedback provided by Tiger.} Our participants mostly felt that the Tiger's feedbacks were appropriate in terms of perceptibility and comfortability. Figure \ref{fig:feedback_rating} shows the distribution of scores for the 7-point Likert scale questionnaires. Feedback 1 means the first vibration feedback provided when a user views a screen for 20 minutes. Feedback 2 is the flashing light feedback provided when taking a break. Feedback 3 is the vibration feedback after a 20-second break. Most of them reported that the three kinds of feedback are well perceptible and comfortable. 

Most of the participants agreed that the strength, length and interval of vibration for the first and third feedback were appropriate to perceive and they were comfortable. Also, they answered that they could easily distinguish the two vibration patterns. P6 said, \textit{"I recognized the vibrations well. I felt the first vibration after 20-minute screen viewing was a bit strong and it was good to notice (...) I felt the other vibration after a break was a bit weak, which was good (...) The third feedback was easily distinguishable from the first feedback.}

Regarding the LED, most of the participants were satisfied with its brightness and flashing interval. However, some reported that they felt uncomfortable with the light. P8 said, \textit{"It was uncomfortable because the LED was too close. I felt the flashing speed was fast and the light was a bit strong."}

In addition, the participants were mostly positive about the functionality of the second feedback to guide the viewing distance during a break. P8 mentioned, "If it simply gave a notification to take a break once, I might end up seeing my smartphone." P3 said, \textit{"I tried to move my head to make green light flash."} On the other hand, P5 reported that the functionality was good, but it was not a must.

\section{Limitations and Discussion}
\label{sec:discussion}

\textbf{User study participants.} The participants of Tiger User Study were students in their 20s. While they do not represent the general public well, we believe that such users would be the main target segment of Tiger as they use screens for a long time and have concerns about their eye health. In this regard, it is meaningful that such a user group feels positive about Tiger. However, due to the characteristics of them, they might be more positive than other general users. To investigate responses from diverse user segments, additional studies with more diverse participants will be necessary.

\textbf{Long-term user study.} We acknowledge that our user study is based on the short-term use of Tiger. In this paper, we mainly focused on studying the design requirements, the techniques for screen viewing detection, and the feedback design to realize an eyewear system for the 20-20-20 rule. To evaluate the effectiveness in real-life scenarios, a long-term user study is necessary. For the purpose, Tiger needs to be improved for everyday use considering wearability.

\textbf{Performance under different circumstances.} The performance of the screen viewing detection algorithm may vary depending on various factors and their combinations, e.g., user mobility, head and hand movements, etc. Since our idea behind the experiment design was that it is crucial to first understand the effect of main factors independently, the current results are limited in terms of covering diverse factors. A rigorous evaluation covering diverse factors and their combinations would be an important future work.

\textbf{Persuasion for behavior change.}
The current feedback by Tiger mainly provides notification upon a relevant event, e.g., vibration after 20 minutes of screen viewing. While our user study shows the feedback is helpful, users might ignore it or stop wearing Tiger when they are repeatedly exposed due to user habituation \cite{anderson2015polymorphic}. A potential solution is to leverage social relationships for behavior change, e.g., social reminders \cite{chiu2009playful}, relational norm intervention \cite{shin2016beupright}.

\textbf{Energy cost.} In this paper, we do not address an energy issue of Tiger prototype. 
However, to get some insights for the room of optimization, we measure the energy cost of the current prototype and break down it into three states; base, interface (for the lidar sensor), and sensing. First, the \textit{base} represents the idle state when no running task on Raspberry Pi Zero W. Its average power is 0.55 W, which is quite high even compared to modern smartphones. This is because Raspberry Pi does not support the low power mode (LPM) yet and we believe a significant energy saving can be achieved if LPM is supported later. Second, we measure the power cost of the UART TTL converter, the \textit{interface} for the lidar sensor. Although the lidar is not activated, the interface consumes 0.55 W. Further energy saving is expected if an energy-efficient interface is provided. Last, the \textit{sensing} represents the power cost for sensing (sensor hardware) and processing (CPU) and its power is 0.51 W. We currently use a high sampling rate for the accuracy, but we can find the optimal rate to balance the accuracy and the power use.

\textbf{Relaxing effect of longer viewing distance.} For the eye resting, the 20-20-20 rule recommends to see 20-feet away, but does not specify the effect of a particular viewing distance. However, it would be possible to have a greater relaxing effect of longer viewing distance. It can be easily incorporated into Tiger if necessary clinical validations become available as Tiger is capable of detecting distances greater than 20 feet with reasonable accuracy.
\section{Conclusion}
We present Tiger, an eyewear system to help users follow the 20-20-20 rule to alleviate the CVS symptoms. We design a light-weight technique for screen viewing detection leveraging color, IMU, and lidar sensors. 
Our evaluation shows that Tiger accurately detects screen viewing events, and is robust to the diverse conditions. Our user study shows the positive perception of Tiger about its usefulness and acceptance.

\begin{acks}
This work was in part supported by the National Research Foundation of Korea (NRF) grant funded by the Korea government (No. 2017R1C1B1010619, 2017M3C4A7066473).
\end{acks}

\balance{}
%
\bibliographystyle{ACM-Reference-Format}
\bibliography{sigchi_bib}


\begin{thebibliography}{00}


\ifx \showCODEN    \undefined \def \showCODEN     #1{\unskip}     \fi
\ifx \showDOI      \undefined \def \showDOI       #1{#1}\fi
\ifx \showISBNx    \undefined \def \showISBNx     #1{\unskip}     \fi
\ifx \showISBNxiii \undefined \def \showISBNxiii  #1{\unskip}     \fi
\ifx \showISSN     \undefined \def \showISSN      #1{\unskip}     \fi
\ifx \showLCCN     \undefined \def \showLCCN      #1{\unskip}     \fi
\ifx \shownote     \undefined \def \shownote      #1{#1}          \fi
\ifx \showarticletitle \undefined \def \showarticletitle #1{#1}   \fi
\ifx \showURL      \undefined \def \showURL       {\relax}        \fi
\providecommand\bibfield[2]{#2}
\providecommand\bibinfo[2]{#2}
\providecommand\natexlab[1]{#1}
\providecommand\showeprint[2][]{arXiv:#2}

\bibitem[\protect\citeauthoryear{??}{cao}{[n. d.]}]%
        {cao202020rule}
 \bibinfo{year}{[n. d.]}\natexlab{}.
\newblock \bibinfo{title}{The 20-20-20 Rule}.
\newblock
  \bibinfo{howpublished}{\url{https://opto.ca/health-library/the-20-20-20-rule}}.
    (\bibinfo{year}{[n. d.]}).
\newblock
\newblock
\shownote{Accessed: January 23, 2019.}


\bibitem[\protect\citeauthoryear{??}{Bre}{[n. d.]}]%
        {Breaks}
 \bibinfo{year}{[n. d.]}\natexlab{}.
\newblock \bibinfo{title}{Breaks for Eyes}.
\newblock
  \bibinfo{howpublished}{\url{https://itunes.apple.com/gb/app/breaks-for-eyes-rest-on-time/id1439431081?mt=12}}.
    (\bibinfo{year}{[n. d.]}).
\newblock
\newblock
\shownote{Accessed: January 23, 2019.}


\bibitem[\protect\citeauthoryear{??}{aoa}{[n. d.]}]%
        {aoacvs}
 \bibinfo{year}{[n. d.]}\natexlab{}.
\newblock \bibinfo{title}{Computer Vision Syndrome}.
\newblock
  \bibinfo{howpublished}{\url{https://www.aoa.org/patients-and-public/caring-for-your-vision/protecting-your-vision/computer-vision-syndrome?sso=y}}.
    (\bibinfo{year}{[n. d.]}).
\newblock
\newblock
\shownote{Accessed: January 23, 2019.}


\bibitem[\protect\citeauthoryear{??}{aao}{[n. d.]}]%
        {aaoeyestrain}
 \bibinfo{year}{[n. d.]}\natexlab{}.
\newblock \bibinfo{title}{Computers, Digital Devices and Eye Strain}.
\newblock
  \bibinfo{howpublished}{\url{https://www.aao.org/eye-health/tips-prevention/computer-usage}}.
    (\bibinfo{year}{[n. d.]}).
\newblock
\newblock
\shownote{Accessed: January 23, 2019.}


\bibitem[\protect\citeauthoryear{??}{eye}{[n. d.]}]%
        {eyecare}
 \bibinfo{year}{[n. d.]}\natexlab{}.
\newblock \bibinfo{title}{Eye Care 20 20 20}.
\newblock
  \bibinfo{howpublished}{\url{https://itunes.apple.com/us/app/eye-care-20-20-20/id967901219?mt=8}}.
    (\bibinfo{year}{[n. d.]}).
\newblock
\newblock
\shownote{Accessed: January 23, 2019.}


\bibitem[\protect\citeauthoryear{??}{Eye}{[n. d.]}]%
        {EyeLeo}
 \bibinfo{year}{[n. d.]}\natexlab{}.
\newblock \bibinfo{title}{EyeLeo}.
\newblock \bibinfo{howpublished}{\url{http://eyeleo.com}}.
  (\bibinfo{year}{[n. d.]}).
\newblock
\newblock
\shownote{Accessed: January 23, 2019.}


\bibitem[\protect\citeauthoryear{??}{eye}{2015}]%
        {eyerestnotification}
 \bibinfo{year}{2015}\natexlab{}.
\newblock \bibinfo{title}{Eye Rest Notification}.
\newblock
  \bibinfo{howpublished}{\url{https://chrome.google.com/webstore/detail/eye-rest-notification/jkijdhebcbpgplameoaifiimkbmlpmeo?hl=en}}.
    (\bibinfo{year}{2015}).
\newblock
\newblock
\shownote{Accessed: January 23, 2019.}


\bibitem[\protect\citeauthoryear{Anderson, Kirwan, Jenkins, Eargle, Howard, and
  Vance}{Anderson et~al\mbox{.}}{2015}]%
        {anderson2015polymorphic}
\bibfield{author}{\bibinfo{person}{Bonnie~Brinton Anderson},
  \bibinfo{person}{C~Brock Kirwan}, \bibinfo{person}{Jeffrey~L Jenkins},
  \bibinfo{person}{David Eargle}, \bibinfo{person}{Seth Howard}, {and}
  \bibinfo{person}{Anthony Vance}.} \bibinfo{year}{2015}\natexlab{}.
\newblock \showarticletitle{How polymorphic warnings reduce habituation in the
  brain: Insights from an fMRI study}. In \bibinfo{booktitle}{{\em Proceedings
  of CHI '15}}. \bibinfo{publisher}{ACM}, \bibinfo{pages}{2883--2892}.
\newblock


\bibitem[\protect\citeauthoryear{Cauchard, L{\"o}chtefeld, Irani, Schoening,
  Kr{\"u}ger, Fraser, and Subramanian}{Cauchard et~al\mbox{.}}{2011}]%
        {cauchard2011visual}
\bibfield{author}{\bibinfo{person}{Jessica~R Cauchard}, \bibinfo{person}{Markus
  L{\"o}chtefeld}, \bibinfo{person}{Pourang Irani}, \bibinfo{person}{Johannes
  Schoening}, \bibinfo{person}{Antonio Kr{\"u}ger}, \bibinfo{person}{Mike
  Fraser}, {and} \bibinfo{person}{Sriram Subramanian}.}
  \bibinfo{year}{2011}\natexlab{}.
\newblock \showarticletitle{Visual separation in mobile multi-display
  environments}. In \bibinfo{booktitle}{{\em Proceedings of the 24th annual ACM
  symposium on User interface software and technology}}. ACM,
  \bibinfo{pages}{451--460}.
\newblock


\bibitem[\protect\citeauthoryear{Chiu, Chang, Chang, Chu, Chen, Hsiao, and
  Ko}{Chiu et~al\mbox{.}}{2009}]%
        {chiu2009playful}
\bibfield{author}{\bibinfo{person}{Meng-Chieh Chiu}, \bibinfo{person}{Shih-Ping
  Chang}, \bibinfo{person}{Yu-Chen Chang}, \bibinfo{person}{Hao-Hua Chu},
  \bibinfo{person}{Cheryl Chia-Hui Chen}, \bibinfo{person}{Fei-Hsiu Hsiao},
  {and} \bibinfo{person}{Ju-Chun Ko}.} \bibinfo{year}{2009}\natexlab{}.
\newblock \showarticletitle{Playful bottle: a mobile social persuasion system
  to motivate healthy water intake}. In \bibinfo{booktitle}{{\em Proceedings of
  UbiComp '09}}. \bibinfo{publisher}{ACM}, \bibinfo{pages}{185--194}.
\newblock


\bibitem[\protect\citeauthoryear{Council}{Council}{2016}]%
        {2016digital}
\bibfield{author}{\bibinfo{person}{The~Vision Council}.}
  \bibinfo{year}{2016}\natexlab{}.
\newblock \bibinfo{title}{2016 Digital Eye Strain Report}.
\newblock
  \bibinfo{howpublished}{\url{https://visionimpactinstitute.org/wp-content/uploads/2016/03/2016EyeStrain_Report_WEB.pdf}}.
    (\bibinfo{year}{2016}).
\newblock
\newblock
\shownote{Accessed: January 23, 2019.}


\bibitem[\protect\citeauthoryear{Crnovrsanin, Wang, and Ma}{Crnovrsanin
  et~al\mbox{.}}{2014}]%
        {crnovrsanin2014stimulating}
\bibfield{author}{\bibinfo{person}{Tarik Crnovrsanin}, \bibinfo{person}{Yang
  Wang}, {and} \bibinfo{person}{Kwan-Liu Ma}.} \bibinfo{year}{2014}\natexlab{}.
\newblock \showarticletitle{Stimulating a Blink: Reduction of Eye Fatigue with
  Visual Stimulus}. In \bibinfo{booktitle}{{\em Proceedings of CHI '14}}.
  \bibinfo{publisher}{ACM}, \bibinfo{pages}{2055--2064}.
\newblock
\showISBNx{978-1-4503-2473-1}


\bibitem[\protect\citeauthoryear{Delamare, Han, and Irani}{Delamare
  et~al\mbox{.}}{2017}]%
        {delamare2017designing}
\bibfield{author}{\bibinfo{person}{William Delamare}, \bibinfo{person}{Teng
  Han}, {and} \bibinfo{person}{Pourang Irani}.}
  \bibinfo{year}{2017}\natexlab{}.
\newblock \showarticletitle{Designing a gaze gesture guiding system}. In
  \bibinfo{booktitle}{{\em Proceedings of the 19th International Conference on
  Human-Computer Interaction with Mobile Devices and Services}}. ACM,
  \bibinfo{pages}{26}.
\newblock


\bibitem[\protect\citeauthoryear{Dementyev and Holz}{Dementyev and
  Holz}{2017}]%
        {dementyev2017dualblink}
\bibfield{author}{\bibinfo{person}{Artem Dementyev} {and}
  \bibinfo{person}{Christian Holz}.} \bibinfo{year}{2017}\natexlab{}.
\newblock \showarticletitle{DualBlink: A Wearable Device to Continuously
  Detect, Track, and Actuate Blinking For Alleviating Dry Eyes and Computer
  Vision Syndrome}.
\newblock \bibinfo{journal}{{\em Proc. ACM Interact. Mob. Wearable Ubiquitous
  Technol.\/}} \bibinfo{volume}{1}, \bibinfo{number}{1}, Article
  \bibinfo{articleno}{1} (\bibinfo{date}{March} \bibinfo{year}{2017}),
  \bibinfo{numpages}{19}~pages.
\newblock
\showISSN{2474-9567}


\bibitem[\protect\citeauthoryear{Duchowski}{Duchowski}{2002}]%
        {duchowski2002breadth}
\bibfield{author}{\bibinfo{person}{Andrew~T Duchowski}.}
  \bibinfo{year}{2002}\natexlab{}.
\newblock \showarticletitle{A breadth-first survey of eye-tracking
  applications}.
\newblock \bibinfo{journal}{{\em Behavior Research Methods, Instruments, \&
  Computers\/}} \bibinfo{volume}{34}, \bibinfo{number}{4}
  (\bibinfo{year}{2002}), \bibinfo{pages}{455--470}.
\newblock


\bibitem[\protect\citeauthoryear{Figo, Diniz, Ferreira, and Cardoso}{Figo
  et~al\mbox{.}}{2010}]%
        {figo2010preprocessing}
\bibfield{author}{\bibinfo{person}{Davide Figo}, \bibinfo{person}{Pedro~C
  Diniz}, \bibinfo{person}{Diogo~R Ferreira}, {and} \bibinfo{person}{Jo{\~a}o~M
  Cardoso}.} \bibinfo{year}{2010}\natexlab{}.
\newblock \showarticletitle{Preprocessing techniques for context recognition
  from accelerometer data}.
\newblock \bibinfo{journal}{{\em Personal and Ubiquitous Computing\/}}
  \bibinfo{volume}{14}, \bibinfo{number}{7} (\bibinfo{year}{2010}),
  \bibinfo{pages}{645--662}.
\newblock


\bibitem[\protect\citeauthoryear{Han, Yang, Kim, and Gerla}{Han
  et~al\mbox{.}}{2012}]%
        {han2012eyeguardian}
\bibfield{author}{\bibinfo{person}{Seongwon Han}, \bibinfo{person}{Sungwon
  Yang}, \bibinfo{person}{Jihyoung Kim}, {and} \bibinfo{person}{Mario Gerla}.}
  \bibinfo{year}{2012}\natexlab{}.
\newblock \showarticletitle{EyeGuardian: A Framework of Eye Tracking and Blink
  Detection for Mobile Device Users}. In \bibinfo{booktitle}{{\em Proceedings
  of HotMobile '12}}. \bibinfo{publisher}{ACM}, Article \bibinfo{articleno}{6},
  \bibinfo{numpages}{6}~pages.
\newblock
\showISBNx{978-1-4503-1207-3}


\bibitem[\protect\citeauthoryear{Ho, Pointner, Shih, Lin, Chen, Tseng, and
  Chen}{Ho et~al\mbox{.}}{2015}]%
        {ho2015eyeprotector}
\bibfield{author}{\bibinfo{person}{Jimmy Ho}, \bibinfo{person}{Reinhard
  Pointner}, \bibinfo{person}{Huai-Chun Shih}, \bibinfo{person}{Yu-Chih Lin},
  \bibinfo{person}{Hsuan-Yu Chen}, \bibinfo{person}{Wei-Luan Tseng}, {and}
  \bibinfo{person}{Mike~Y. Chen}.} \bibinfo{year}{2015}\natexlab{}.
\newblock \showarticletitle{EyeProtector: Encouraging a Healthy Viewing
  Distance when Using Smartphones}. In \bibinfo{booktitle}{{\em Proceedings of
  MobileHCI '15}}. \bibinfo{publisher}{ACM}, \bibinfo{pages}{77--85}.
\newblock
\showISBNx{978-1-4503-3652-9}


\bibitem[\protect\citeauthoryear{Kaaresoja and Linjama}{Kaaresoja and
  Linjama}{2005}]%
        {kaaresoja2005perception}
\bibfield{author}{\bibinfo{person}{Topi Kaaresoja} {and} \bibinfo{person}{Jukka
  Linjama}.} \bibinfo{year}{2005}\natexlab{}.
\newblock \showarticletitle{Perception of short tactile pulses generated by a
  vibration motor in a mobile phone}. In \bibinfo{booktitle}{{\em First Joint
  Eurohaptics Conference and Symposium on Haptic Interfaces for Virtual
  Environment and Teleoperator Systems. World Haptics Conference}}.
  \bibinfo{publisher}{IEEE}, \bibinfo{pages}{471--472}.
\newblock


\bibitem[\protect\citeauthoryear{Khamis, Alt, and Bulling}{Khamis
  et~al\mbox{.}}{2018}]%
        {khamis2018past}
\bibfield{author}{\bibinfo{person}{Mohamed Khamis}, \bibinfo{person}{Florian
  Alt}, {and} \bibinfo{person}{Andreas Bulling}.}
  \bibinfo{year}{2018}\natexlab{}.
\newblock \showarticletitle{The past, present, and future of gaze-enabled
  handheld mobile devices: survey and lessons learned}. In
  \bibinfo{booktitle}{{\em Proceedings of the 20th International Conference on
  Human-Computer Interaction with Mobile Devices and Services}}. ACM,
  \bibinfo{pages}{38}.
\newblock


\bibitem[\protect\citeauthoryear{Lee, Min, and Kang}{Lee et~al\mbox{.}}{2018}]%
        {lee2018towards}
\bibfield{author}{\bibinfo{person}{Euihyeok Lee}, \bibinfo{person}{Chulhong
  Min}, {and} \bibinfo{person}{Seungwoo Kang}.}
  \bibinfo{year}{2018}\natexlab{}.
\newblock \showarticletitle{Towards a Wearable Assistant to Prevent Computer
  Vision Syndrome}. In \bibinfo{booktitle}{{\em Proceedings of the 2018 ACM
  International Joint Conference and 2018 International Symposium on Pervasive
  and Ubiquitous Computing and Wearable Computers}}. ACM,
  \bibinfo{pages}{122--125}.
\newblock


\bibitem[\protect\citeauthoryear{Long, Rosenfield, Helland, and Anshel}{Long
  et~al\mbox{.}}{2014}]%
        {long2014visual}
\bibfield{author}{\bibinfo{person}{Jennifer Long}, \bibinfo{person}{Mark
  Rosenfield}, \bibinfo{person}{Magne Helland}, {and} \bibinfo{person}{Jeffrey
  Anshel}.} \bibinfo{year}{2014}\natexlab{}.
\newblock \showarticletitle{Visual ergonomics standards for contemporary office
  environments}.
\newblock \bibinfo{journal}{{\em Ergonomics Australia\/}} \bibinfo{volume}{10},
  \bibinfo{number}{1} (\bibinfo{year}{2014}), \bibinfo{pages}{7}.
\newblock


\bibitem[\protect\citeauthoryear{Mayberry, Hu, Marlin, Salthouse, and
  Ganesan}{Mayberry et~al\mbox{.}}{2014}]%
        {mayberry2014ishadow}
\bibfield{author}{\bibinfo{person}{Addison Mayberry}, \bibinfo{person}{Pan Hu},
  \bibinfo{person}{Benjamin Marlin}, \bibinfo{person}{Christopher Salthouse},
  {and} \bibinfo{person}{Deepak Ganesan}.} \bibinfo{year}{2014}\natexlab{}.
\newblock \showarticletitle{iShadow: Design of a Wearable, Real-time Mobile
  Gaze Tracker}. In \bibinfo{booktitle}{{\em Proceedings of MobiSys '14}}.
  \bibinfo{publisher}{ACM}, \bibinfo{pages}{82--94}.
\newblock
\showISBNx{978-1-4503-2793-0}


\bibitem[\protect\citeauthoryear{Mayberry, Tun, Hu, Smith-Freedman, Ganesan,
  Marlin, and Salthouse}{Mayberry et~al\mbox{.}}{2015}]%
        {mayberry2015cider}
\bibfield{author}{\bibinfo{person}{Addison Mayberry}, \bibinfo{person}{Yamin
  Tun}, \bibinfo{person}{Pan Hu}, \bibinfo{person}{Duncan Smith-Freedman},
  \bibinfo{person}{Deepak Ganesan}, \bibinfo{person}{Benjamin~M. Marlin}, {and}
  \bibinfo{person}{Christopher Salthouse}.} \bibinfo{year}{2015}\natexlab{}.
\newblock \showarticletitle{CIDER: Enabling Robustness-Power Tradeoffs on a
  Computational Eyeglass}. In \bibinfo{booktitle}{{\em Proceedings of MobiCom
  '15}}. \bibinfo{publisher}{ACM}, \bibinfo{pages}{400--412}.
\newblock
\showISBNx{978-1-4503-3619-2}


\bibitem[\protect\citeauthoryear{Nukarinen, Rantala, Farooq, and
  Raisamo}{Nukarinen et~al\mbox{.}}{2015}]%
        {nukarinen2015delivering}
\bibfield{author}{\bibinfo{person}{Tomi Nukarinen}, \bibinfo{person}{Jussi
  Rantala}, \bibinfo{person}{Ahmed Farooq}, {and} \bibinfo{person}{Roope
  Raisamo}.} \bibinfo{year}{2015}\natexlab{}.
\newblock \showarticletitle{Delivering directional haptic cues through
  eyeglasses and a seat}. In \bibinfo{booktitle}{{\em 2015 IEEE World Haptics
  Conference (WHC)}}. \bibinfo{publisher}{IEEE}, \bibinfo{pages}{345--350}.
\newblock


\bibitem[\protect\citeauthoryear{Rantala, Kangas, Akkil, Isokoski, and
  Raisamo}{Rantala et~al\mbox{.}}{2014}]%
        {rantala2014glasses}
\bibfield{author}{\bibinfo{person}{Jussi Rantala}, \bibinfo{person}{Jari
  Kangas}, \bibinfo{person}{Deepak Akkil}, \bibinfo{person}{Poika Isokoski},
  {and} \bibinfo{person}{Roope Raisamo}.} \bibinfo{year}{2014}\natexlab{}.
\newblock \showarticletitle{Glasses with haptic feedback of gaze gestures}. In
  \bibinfo{booktitle}{{\em CHI'14 Extended Abstracts}}.
  \bibinfo{publisher}{ACM}, \bibinfo{pages}{1597--1602}.
\newblock


\bibitem[\protect\citeauthoryear{Rosenfield}{Rosenfield}{[n. d.]}]%
        {mark2011copmuter}
\bibfield{author}{\bibinfo{person}{Mark Rosenfield}.} \bibinfo{year}{[n.
  d.]}\natexlab{}.
\newblock \showarticletitle{Computer vision syndrome: a review of ocular causes
  and potential treatments}.
\newblock \bibinfo{journal}{{\em Ophthalmic and Physiological Optics\/}}
  \bibinfo{volume}{31}, \bibinfo{number}{5} (\bibinfo{year}{[n. d.]}),
  \bibinfo{pages}{502--515}.
\newblock


\bibitem[\protect\citeauthoryear{Saket, Prasojo, Huang, and Zhao}{Saket
  et~al\mbox{.}}{2013}]%
        {saket2013designing}
\bibfield{author}{\bibinfo{person}{Bahador Saket}, \bibinfo{person}{Chrisnawan
  Prasojo}, \bibinfo{person}{Yongfeng Huang}, {and} \bibinfo{person}{Shengdong
  Zhao}.} \bibinfo{year}{2013}\natexlab{}.
\newblock \showarticletitle{Designing an effective vibration-based notification
  interface for mobile phones}. In \bibinfo{booktitle}{{\em Proceedings of CSCW
  '13}}. \bibinfo{publisher}{ACM}, \bibinfo{pages}{1499--1504}.
\newblock


\bibitem[\protect\citeauthoryear{Shin, Kang, Park, Huh, Kim, and Song}{Shin
  et~al\mbox{.}}{2016}]%
        {shin2016beupright}
\bibfield{author}{\bibinfo{person}{Jaemyung Shin}, \bibinfo{person}{Bumsoo
  Kang}, \bibinfo{person}{Taiwoo Park}, \bibinfo{person}{Jina Huh},
  \bibinfo{person}{Jinhan Kim}, {and} \bibinfo{person}{Junehwa Song}.}
  \bibinfo{year}{2016}\natexlab{}.
\newblock \showarticletitle{BeUpright: Posture Correction Using Relational Norm
  Intervention}. In \bibinfo{booktitle}{{\em Proceedings of CHI '16}}.
  \bibinfo{publisher}{ACM}, \bibinfo{pages}{6040--6052}.
\newblock


\bibitem[\protect\citeauthoryear{Vaitukaitis and Bulling}{Vaitukaitis and
  Bulling}{2012}]%
        {vaitukaitis2012eye}
\bibfield{author}{\bibinfo{person}{Vytautas Vaitukaitis} {and}
  \bibinfo{person}{Andreas Bulling}.} \bibinfo{year}{2012}\natexlab{}.
\newblock \showarticletitle{Eye gesture recognition on portable devices}. In
  \bibinfo{booktitle}{{\em Proceedings of the 2012 ACM Conference on Ubiquitous
  Computing}}. ACM, \bibinfo{pages}{711--714}.
\newblock


\bibitem[\protect\citeauthoryear{Wahl, Kasbauer, and Amft}{Wahl
  et~al\mbox{.}}{2017}]%
        {wahl2017computer}
\bibfield{author}{\bibinfo{person}{Florian Wahl}, \bibinfo{person}{Jakob
  Kasbauer}, {and} \bibinfo{person}{Oliver Amft}.}
  \bibinfo{year}{2017}\natexlab{}.
\newblock \showarticletitle{Computer Screen Use Detection Using Smart
  Eyeglasses}.
\newblock \bibinfo{journal}{{\em Frontiers in ICT\/}}  \bibinfo{volume}{4}
  (\bibinfo{year}{2017}), \bibinfo{pages}{8}.
\newblock
\showISSN{2297-198X}


\bibitem[\protect\citeauthoryear{Xiong, Shen, Yang, Lee, and Wu}{Xiong
  et~al\mbox{.}}{2018}]%
        {xiong2018color}
\bibfield{author}{\bibinfo{person}{Neal~N Xiong}, \bibinfo{person}{Yang Shen},
  \bibinfo{person}{Kangye Yang}, \bibinfo{person}{Changhoon Lee}, {and}
  \bibinfo{person}{Chunxue Wu}.} \bibinfo{year}{2018}\natexlab{}.
\newblock \showarticletitle{Color sensors and their applications based on
  real-time color image segmentation for cyber physical systems}.
\newblock \bibinfo{journal}{{\em EURASIP Journal on Image and Video
  Processing\/}} \bibinfo{volume}{2018}, \bibinfo{number}{1}
  (\bibinfo{year}{2018}), \bibinfo{pages}{23}.
\newblock


\bibitem[\protect\citeauthoryear{Yan, Hu, Chen, and Lu}{Yan
  et~al\mbox{.}}{2008}]%
        {yan2008computer}
\bibfield{author}{\bibinfo{person}{Zheng Yan}, \bibinfo{person}{Liang Hu},
  \bibinfo{person}{Hao Chen}, {and} \bibinfo{person}{Fan Lu}.}
  \bibinfo{year}{2008}\natexlab{}.
\newblock \showarticletitle{Computer Vision Syndrome: A widely spreading but
  largely unknown epidemic among computer users}.
\newblock \bibinfo{journal}{{\em Computers in Human Behavior\/}}
  \bibinfo{volume}{24}, \bibinfo{number}{5} (\bibinfo{year}{2008}),
  \bibinfo{pages}{2026 -- 2042}.
\newblock
\showISSN{0747-5632}


\bibitem[\protect\citeauthoryear{Zhang, Li, Huang, Liu, Zong, Jian, Feng, Jung,
  and Liu}{Zhang et~al\mbox{.}}{2014}]%
        {zhang2014starts}
\bibfield{author}{\bibinfo{person}{Lan Zhang}, \bibinfo{person}{Xiang-Yang Li},
  \bibinfo{person}{Wenchao Huang}, \bibinfo{person}{Kebin Liu},
  \bibinfo{person}{Shuwei Zong}, \bibinfo{person}{Xuesi Jian},
  \bibinfo{person}{Puchun Feng}, \bibinfo{person}{Taeho Jung}, {and}
  \bibinfo{person}{Yunhao Liu}.} \bibinfo{year}{2014}\natexlab{}.
\newblock \showarticletitle{It Starts with iGaze: Visual Attention Driven
  Networking with Smart Glasses}. In \bibinfo{booktitle}{{\em Proceedings of
  MobiCom '14}}. \bibinfo{publisher}{ACM}, \bibinfo{pages}{91--102}.
\newblock
\showISBNx{978-1-4503-2783-1}


\bibitem[\protect\citeauthoryear{Zhang and Rehg}{Zhang and Rehg}{2018}]%
        {zhang2018watching}
\bibfield{author}{\bibinfo{person}{Yun~C. Zhang} {and}
  \bibinfo{person}{James~M. Rehg}.} \bibinfo{year}{2018}\natexlab{}.
\newblock \showarticletitle{Watching the TV Watchers}.
\newblock \bibinfo{journal}{{\em Proc. ACM Interact. Mob. Wearable Ubiquitous
  Technol.\/}} \bibinfo{volume}{2}, \bibinfo{number}{2}, Article
  \bibinfo{articleno}{88} (\bibinfo{date}{June} \bibinfo{year}{2018}),
  \bibinfo{numpages}{27}~pages.
\newblock
\showISSN{2474-9567}


\end{thebibliography}

%

\end{document}